\documentclass[
 reprint,
superscriptaddress,
 amsmath,amssymb,
 aps,
pra,
]{revtex4-1}

\usepackage{graphicx}
\usepackage{dcolumn}
\usepackage{bm}
\usepackage{hyperref}
\usepackage{color}
\usepackage{subfigure}
\usepackage{tikz}
\usepackage{epstopdf}
\usepackage{pgfplots}

\begin{document}

\preprint{APS/123-QED}

\title{Design of a Surface Trap for Freely Rotating Ion Ring Crystals}

\author{Po-Jen Wang}
 \affiliation{Department of Physics, University of California, Berkeley, 
California 94720, USA}

 \author{Tongcang Li}
\affiliation{NSF Nanoscale Science and Engineering Center, 3112 Etcheverry Hall, 
University of California, Berkeley, California 94720, USA}

 \author{Crystal Noel}
\affiliation{Department of Physics, University of California, Berkeley, 
California 94720, USA}

 \author{Xiang Zhang}
\email{xiang@berkeley.edu}
\affiliation{NSF Nanoscale Science and Engineering Center, 3112 Etcheverry Hall, 
University of California, Berkeley, California 94720, USA}

\author{Hartmut H\"affner}
\email{hhaeffner@berkeley.edu}
\affiliation{Department of Physics, University of California, Berkeley, 
California 94720, USA}

\date{\today}

\begin{abstract}
 We present a design of an r.f. trap using planar electrodes with the goal to 
trap on the order of 100 ions in a small ring structure of diameters ranging 
between 100~$\mu$m and  200~$\mu$m. In order to minimize the influence of 
 trap electrode imperfections due to the fabrication, we aim at trapping the ions around 400~$\mu$m 
above the trap electrodes. 
 In view of experiments to create freely rotating crystals near the ground 
state, we numerically study factors breaking the rotational symmetry such as external stray electric fields, local charging of the trap 
electrodes, and fabrication imperfections. We conclude that these imperfections can be controlled
sufficiently well under state-of-the-art experimental conditions to allow for freely rotating ion rings even at energies
comparable to the ground state energy of the rotational degree-of-freedom.
\end{abstract}


\maketitle

\section{Introduction}

The electronic and motional degrees-of-freedom of ions trapped with 
electromagnetic fields are extremely well decoupled from their environment. In 
addition, lasers and electromagnetic fields allow for excellent control of both 
degrees-of-freedom on the single quantum level 
\cite{Leibfried2003,Haeffner2008}. Both of those properties make ion crystals 
nearly perfect systems to study many-body physics in closed systems 
\cite{Friedenauer2008a,Richerme2014,Jurcevic2014,Ramm2014}. While most 
experiments are carried out with linear ion strings, a particularly interesting 
structure is a ring of trapped ions. Proposals include mini-accelerators 
\cite{Blumel1999,Schatz2001}, dynamics of Kinks\cite{Landa2010}, quantum 
emulation of ring molecules, and the acoustic analog of Hawking radiation 
\cite{Horstmann2010}. Recently, also rings of trapped ions have been suggested 
to realize the concept of so-called time crystals \cite{Wilczek2012,Li2012}. 

However, starting from experiments on how to implement rings of 
trapped ions by Sch\"atz {\em et. al.} \cite{Waki1992,Schatz2001}, it has become clear 
that imperfections and charging will make it very hard to implement such 
experiments. Thus, a few design improvements have been proposed 
\cite{Lammert2006,Austin2007,Madsen2010}. Furthermore, the Sandia group has 
implemented a ring trap on surface trap technology \cite{Tabakov2012}. Common to 
those designs and experiments is that the resulting ring potential has 
relatively large diameters, making it difficult to compensate for imperfections.

Inspired by Ref.~\cite{Clark2013}, we study a novel design, deviating from the 
idea of bending a conventional linear trap into a ring. In addition, with a 
planar electrode design amenable to microfabrication, we seek to reduce 
inevitable imperfections from the geometry as well as local charging of trap 
electrodes. The main feature of our geometry is to trap the ion ring far away 
from the trapping electrodes  as compared to the typical ion-ion distance and 
the ring diameter itself. Thus local imperfections from stray charges affect the 
rotational symmetry of the ion ring much less as if the ions were trapped close 
to the trap electrodes. 

\begin{figure}[ht!]
			\begin{center}
			\includegraphics[width=2.5in]{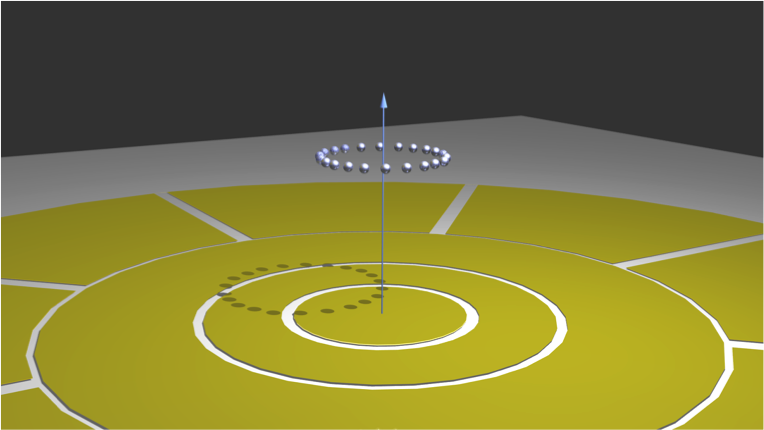}
			\caption{(Color online) Schematic of the trap showing ring-shaped 
surface electrodes in the center. A ring of ions (floating balls in the figure) are trapped
about 400~$\mu$m above the surface.The fan-shaped electrodes are 
compensation electrodes. We neglect gaps between the electrodes 
justified from our experience with linear surface traps. }
			\label{fig:concept}
			\end{center}
			\end{figure}

Our design is composed of concentric planar 
ring electrodes (see Fig.~\ref{fig:concept}). Trapping is accomplished by applying suitable radio frequency 
(rf) voltage to those rings. Work by Clark suggest that the multipoles of such a 
trapping potential can be adjusted over a wide range by changing the rf voltage 
on the various rings \cite{Clark2013}. However, it is experimentally difficult 
to keep several rf high-voltage sources in phase. We thus design a trap 
requiring only one rf high-voltage source. Fixing this parameter, we aim to find 
a trap geometry yielding a rotationally symmetric potential minimum at the 
desired ring diameter and height.

For the study of the physics of symmetrization of the wavefunction 
of bosonic and fermionic ions
 discussed in the context of time 
crystals\cite{Wilczek2012,Li2012}, it will be important that the ion ring can 
freely rotate at the level of single rotational quanta. Keeping that in mind, we 
consider $N$ identical ions with mass $M$
 and charge $q$ in a ring trap with diameter $d$ and a uniform magnetic field. 
The energy scales of its internal vibration modes are much larger than the 
energy scale of its collective rotation \cite{Burmeister2002}. Thus, we focus on the  
collective rotation. Because of the required symmetrization of identical ions, 
the canonical angular momentum $L$ of the ion ring will be quantized according 
to $L=N\hbar n_r$ for identical bosons, where $n_r$ is the quantum number of the 
rotation. The kinetic angular momentum is $K=N\hbar(n_r-\alpha)$  and the 
eigenenergy of the collective rotation is  
$E_{n_r}=\frac{2N\hbar^2}{Md^2}(n_r-\alpha)^2$ for identical bosonic ions, where 
 $\alpha$ is the normalized magnetic flux. This provides an energy scale of 
 \begin{equation}\label{eq:gap}
 E_{{\rm gap}}=\frac{N\hbar^2}{Md^2} \:.
 \end{equation}
 For identical fermions, the energy scale is the same, although its dependence 
on  $\alpha$ can be different. For an ion ring of 100~$^{40}$Ca$^+$ ions with a diameter of 100~$\mu$m, the 
energy scale of the collective rotation is $E_{{\rm gap}}/k_B=0.2$~nK. 
We seek to create a ring potential with 
sufficiently small imperfections such that a classical ion ring with rotational energy 
corresponding to the ground state energy $E_{{\rm gap}}/2$ would not be pinned
by the imperfections.

In view of these considerations, it is important to trap the ion ring far away 
from trapping electrodes, while at the same time keeping the ion ring as compact 
as possible. Thus, heating is reduced while still maintaining a reasonable 
energy scale of the ground state. 
Additional design constraints are ease of symmetric trap fabrication as well as 
reasonable trapping voltages while maintaining an appreciable trap depth.
To this end we target an ion ring with a diameter of 100~$\mu$m trapped 
400~$\mu$m above the trap electrodes. 

The remainder of the paper is organized as follows. In 
Sec.~\ref{sec:how-to-calculate}, we briefly summarize the methods outlined in 
Ref.~\cite{Clark2013} on how to efficiently calculate rotationally symmetric 
potentials. 
Armed with the potential, we study the structure of ion crystals forming in such 
ring shaped potentials in Sec.~\ref{sec:ion-structures}. We then analyze various 
imperfections breaking the rotational symmetry in Sec.~\ref{sec:imperfections}, 
most notably external stray fields, electrode edge irregularities, and local charging of the trap 
electrodes. Sec.~\ref{sec:lasercooling} addresses the process of cooling and pinning such a ring of ions.

			\begin{figure}[ht!]
			\begin{center}
			\includegraphics[width=3in]{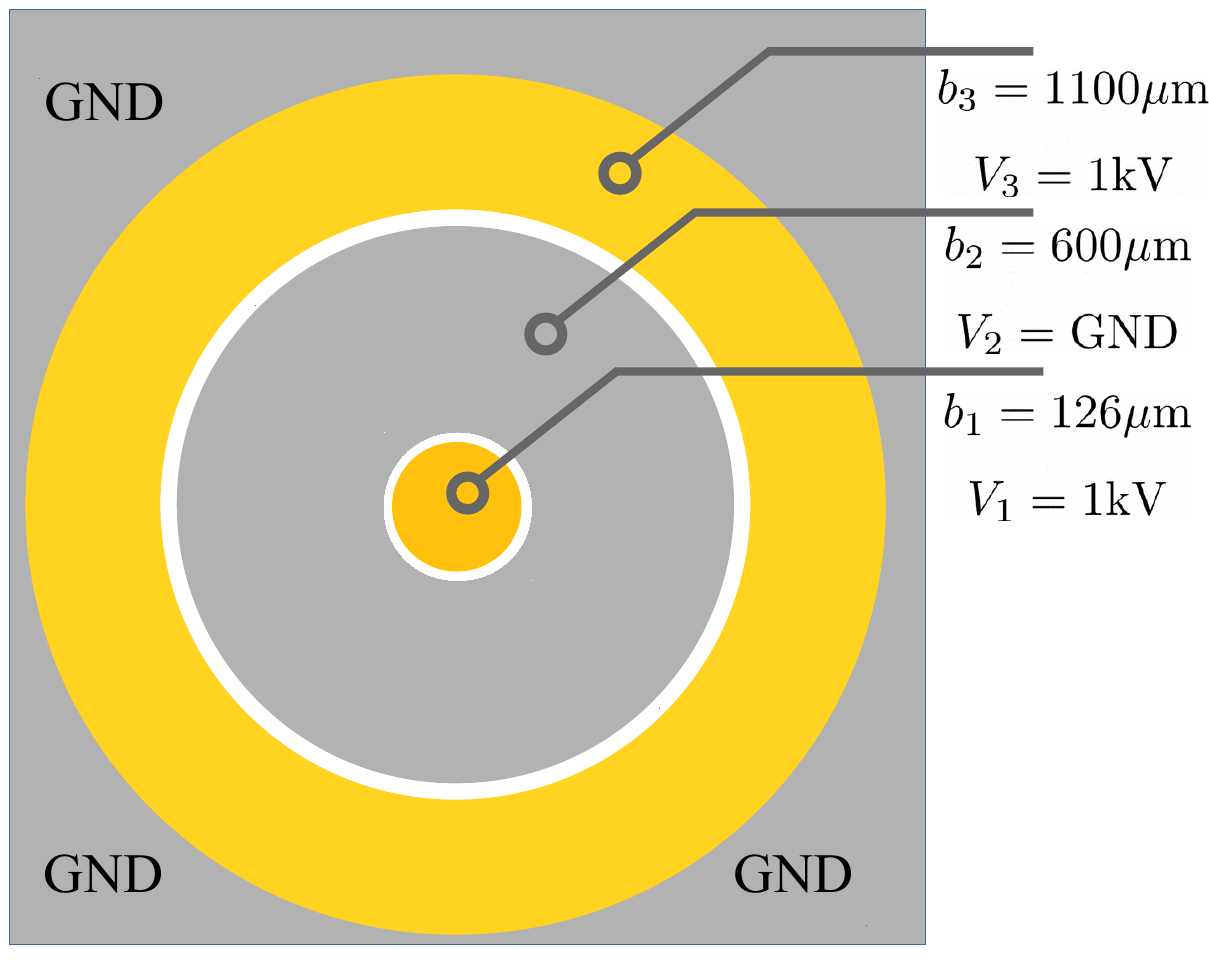}
			\caption{(Color online) Top view of the trap showing details of the 
design, specifying the outer radius $b_i$ and voltage amplitude $V_i$ for each electrode. The surface
trap considered in this paper has three ring electrodes and one large ground electrode that covers all
other parts of the surface. The first(center) and third ring electrodes are connected to the same high voltage RF source. 
The second ring electrode will be connected to ground or a small RF voltage for fine tuning of the diameter of the ring
ion crystal. }
			\label{fig:ring}
			\end{center}
			\end{figure}

\section{Calculation of the trapping potential}
\label{sec:how-to-calculate}
	\paragraph*{}
	We start with the trap design proposed in Ref.\cite{Clark2013} composed 
of planar ring electrodes of different radii and applied voltages. Given this 
cylindrically symmetric boundary condition, the analytic solution to Laplace's 
equation is given by \cite{Kim2010b}
			
	\begin{equation}
	 \Phi(z, r) = \int_0^{\infty} J_0(kr)e^{-kz}A_0(k)dk\:.
	 \end{equation}
where $J_i$ is the Bessel function of $i^{th}$ order. $ A_0(k)$ can be expressed 
as $A_0(k) = \sum^N_{i=1}A_i(k)$, and $A_i(k)$ is given by
	\begin{equation}
	A_i(k) = V_i(b_{i}J_1(kb_{i}) - a_iJ_1(ka_i))\:.
	\end{equation}
where $b_i$ and $a_i$ are the outer and inner radius of each ring electrode and 
$V_i$ is the amplitude of the rf voltage applied to the each electrode.  In 
order to study ions in this oscillating trapping potential, we approximate the 
potential by the time-averaged pseudopotential
	\begin{equation}
	\Psi(z,r) = \frac{Q^2}{4M\Omega^2_{\rm{rf}}}|\vec{E}(z,r)|^2\:.
	\end{equation}
This approximation is valid when the oscillation frequency of the trapped ion is 
much smaller than the rf frequency.  
		
\begin{figure}
			\begin{center}
			\includegraphics[width=3.5in]{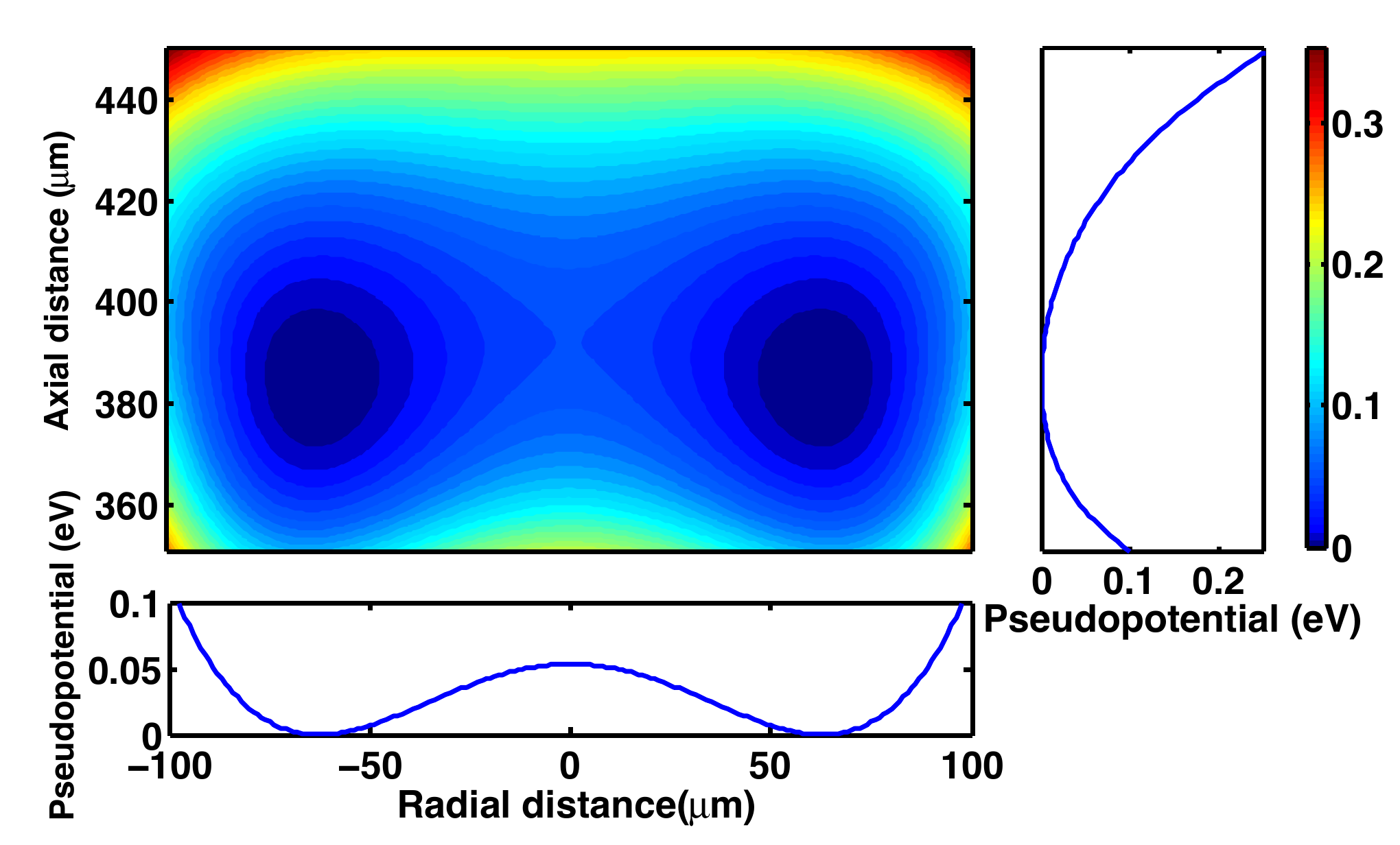}
			\caption{(Color online) Contour plot of the pseudopotential. 
			The lower subfigure is the cross section at the trapping 
height, z = 385~$\mu$m. The right subfigure is the cross section at the radial 
trapping position, r = 58~$\mu$m.}
			\label{fig:pseudop}
			\end{center}
			\end{figure}

	\paragraph*{Trap design}
	 Ref.\cite{Clark2013} showed that multipole surface traps can be built 
from concentric rings with particular sets of ring diameters and applied 
voltages. However, it is experimentally difficult to keep several rf 
high-voltage sources in phase. We thus design a trap targeting only one rf 
high-voltage source. Our design,  shown in Fig.~\ref{fig:ring}, is composed of 
three concentric ring electrodes with outer radius $b_i = \{$126~$\mu$m, 
600~$\mu$m, 1100~$\mu$m$\}$, with the second ring grounded and the other two 
connected to a fixed rf driving source of amplitude $V_{rf}$ = 1000~V. As the rf 
driving frequency, we choose $\Omega= 6$~MHz. In what follows, we also assume 
Calcium ions with mass $M = 40$~amu. The design leads to a Mexican-hat-shaped 
pseudopotential in the radial direction and a confining pseudopotential in the 
axial direction, as shown in Fig.~\ref{fig:pseudop}. The trap potential has 
minimum at radius $r\approx 58\:\mu$m, height $h\approx$ 385~$\mu$m, and a trap 
depth of 0.134~eV, leading to single-ion trap frequencies $f_r = 1.03$~MHz and 
$f_z = 1.02$~MHz in the radial and vertical direction, respectively. 

\begin{figure}[ht!]
			\begin{center}
				\includegraphics[width=3.2in]{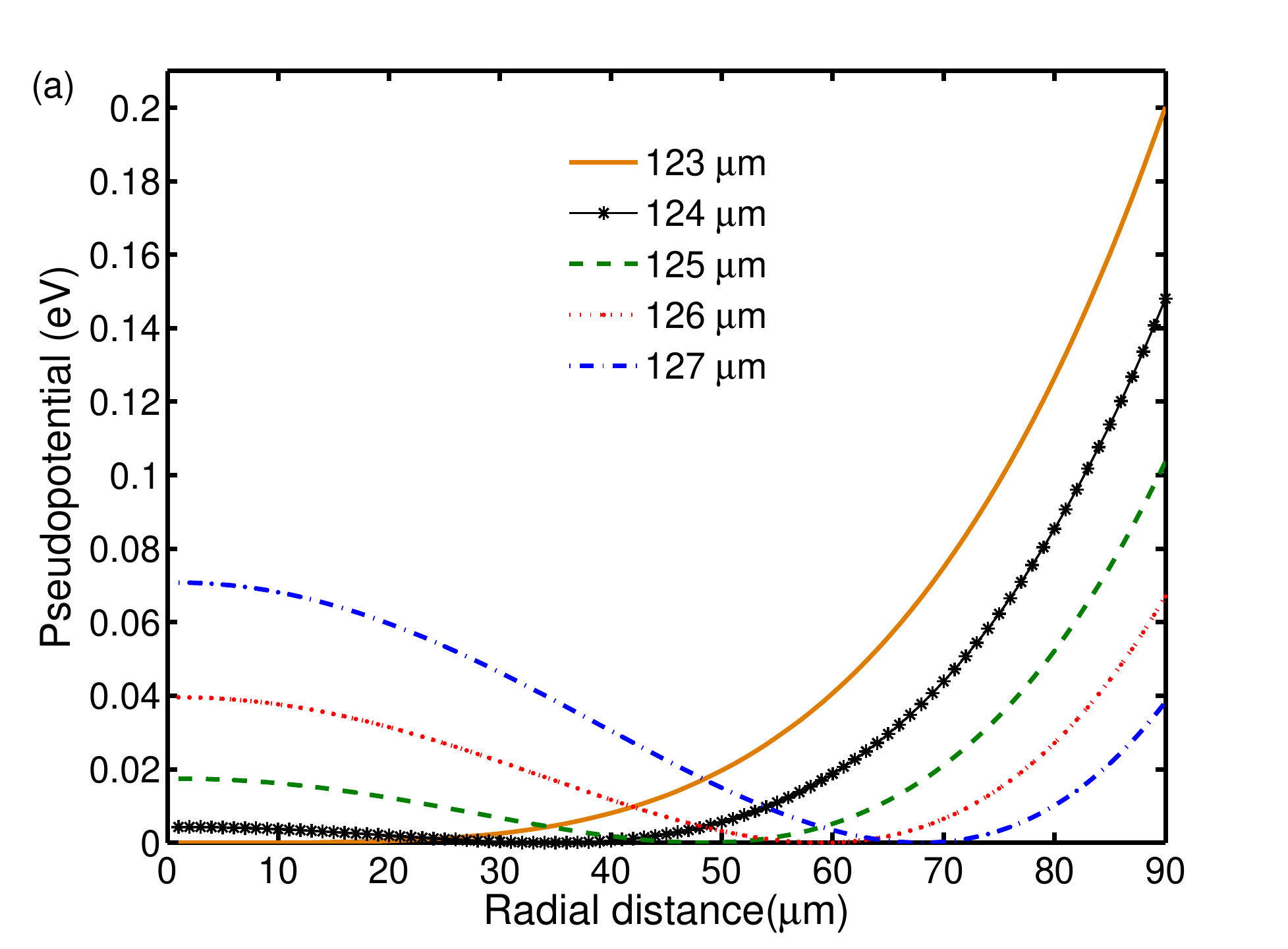}
				\includegraphics[width=3.2in]{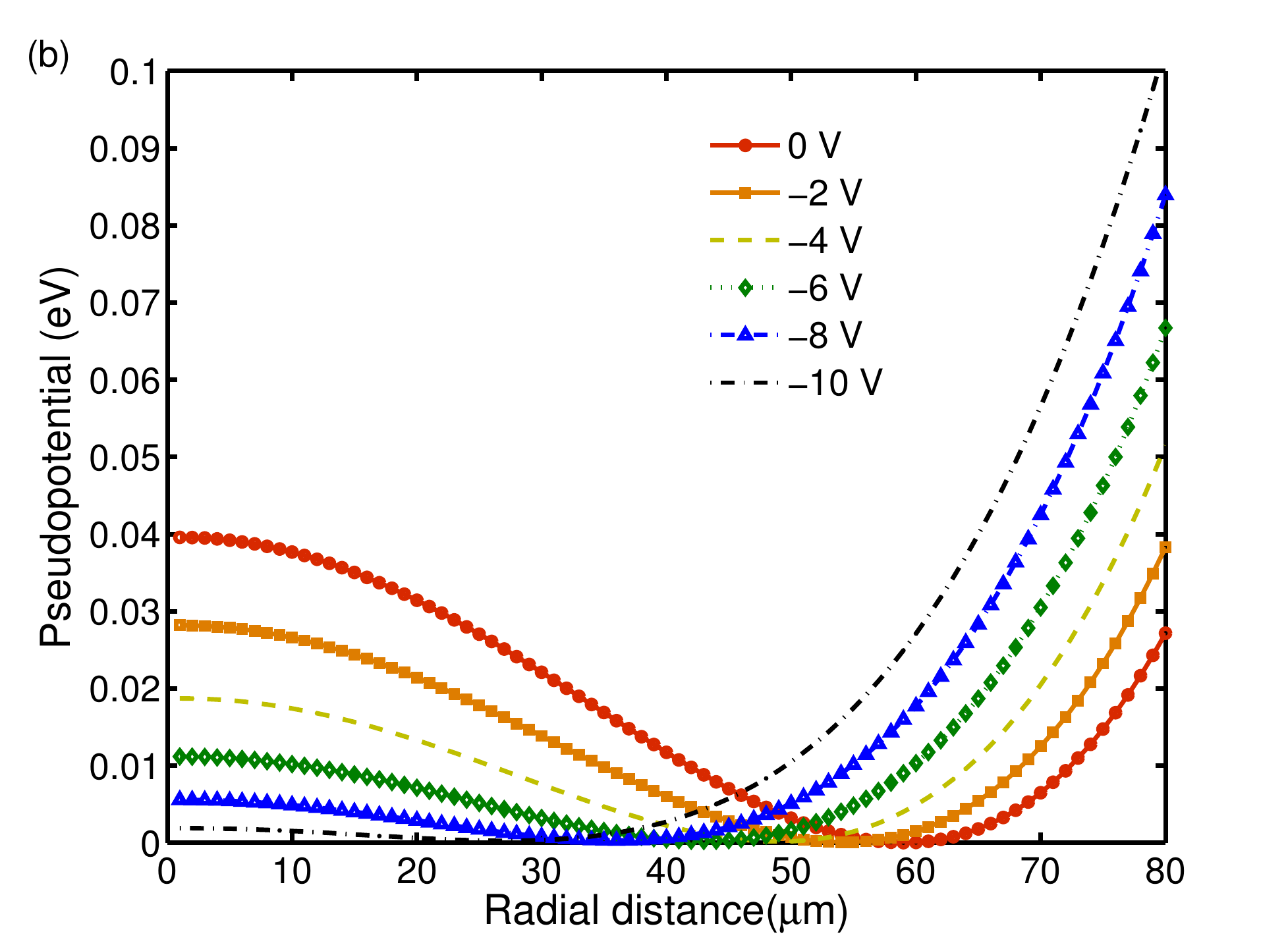}
				\caption{(Color online) (a) The effect on the pseudopotential 
due to varying the size of the center electrode (b) the effect of varying the amplitude
of the compensation RF voltage applied to the second ring electrode.}
			\label{fig:parameter}
			\end{center}
			\end{figure}

	\paragraph*{Design variation compensation}
	In view of fabrication imperfections, we study the effects of small 
deviation in the size of the center electrode. Our simulation 
Fig.~\ref{fig:parameter} shows that the position of the potential minimum is 
very sensitive to the change in the size of the center electrode. In particular, 
it shows that changes of 1~$\mu$m in radius will shift the radius of the minimum 
by 10~$\mu$m. While we expect that microfabrication allows fabrication with 
tolerances below the micrometer range, we also can tune the potential by adding a 
small variable rf voltage  with the same driving frequency  on the second ring, 
but with the phase exactly opposite to that. Simulations show that the minimum 
position is shifted radially inward by about 2.5~$\mu$m/V, while the trap depth 
changes by 0.005~eV/V. This small compensation rf voltage provides a powerful 
tool for fine-tuning the potential {\em in situ}.
	
\section{Structure of ring crystals}
\label{sec:ion-structures}
	\paragraph*{}
		Of particular interest are the conditions under which ultra-cold ions 
form a ring in this potential. For this, we carry out molecular dynamics 
simulation to analyze the structure of laser-cooled ions in the surface trap 
\cite{Okada2007}. We calculate the trajectories and velocities of the trapped 
ions by solving Newtonian equations of motion including the Coulomb interaction, 
the pseudoforce from the rf potential, and a hypothetical damping force term 
\cite{Meyrath1998}. The extra damping term serves as a friction term that will 
gradually reduce the energy of the ions, thereby simulating laser cooling. Thus, 
the ions will eventually reach a steady state, which represents the expected 
structure of the cold ion crystal. The equations of motion of the 
$\textit{i}^{\rm th}$ ion can be written
		\begin{equation}
			m_i\frac{d^2\vec{x}_i}{dt^2} = - 
\gamma\frac{d\vec{x}_i}{dt} +  \vec{F_{T}} + \vec{F_{C}}\:.
		\end{equation}
		where $\gamma$ is the damping coefficient, $\vec{F_{T}}$ is the 
pseudoforce and the Coulomb force $\vec{F_{C}}$ is given by
		\begin{equation}
			\vec{F_{C}} = 
\frac{q^2}{4\pi\epsilon_0}\sum\limits_{i\neq 
j}^N\frac{\vec{r_i}-\vec{r_j}}{R_{ij}^3}\:,
		\end{equation}
		where $N$ is the number of ions in the trap.

\begin{figure*}
			\begin{center}
			\includegraphics[width=5in]{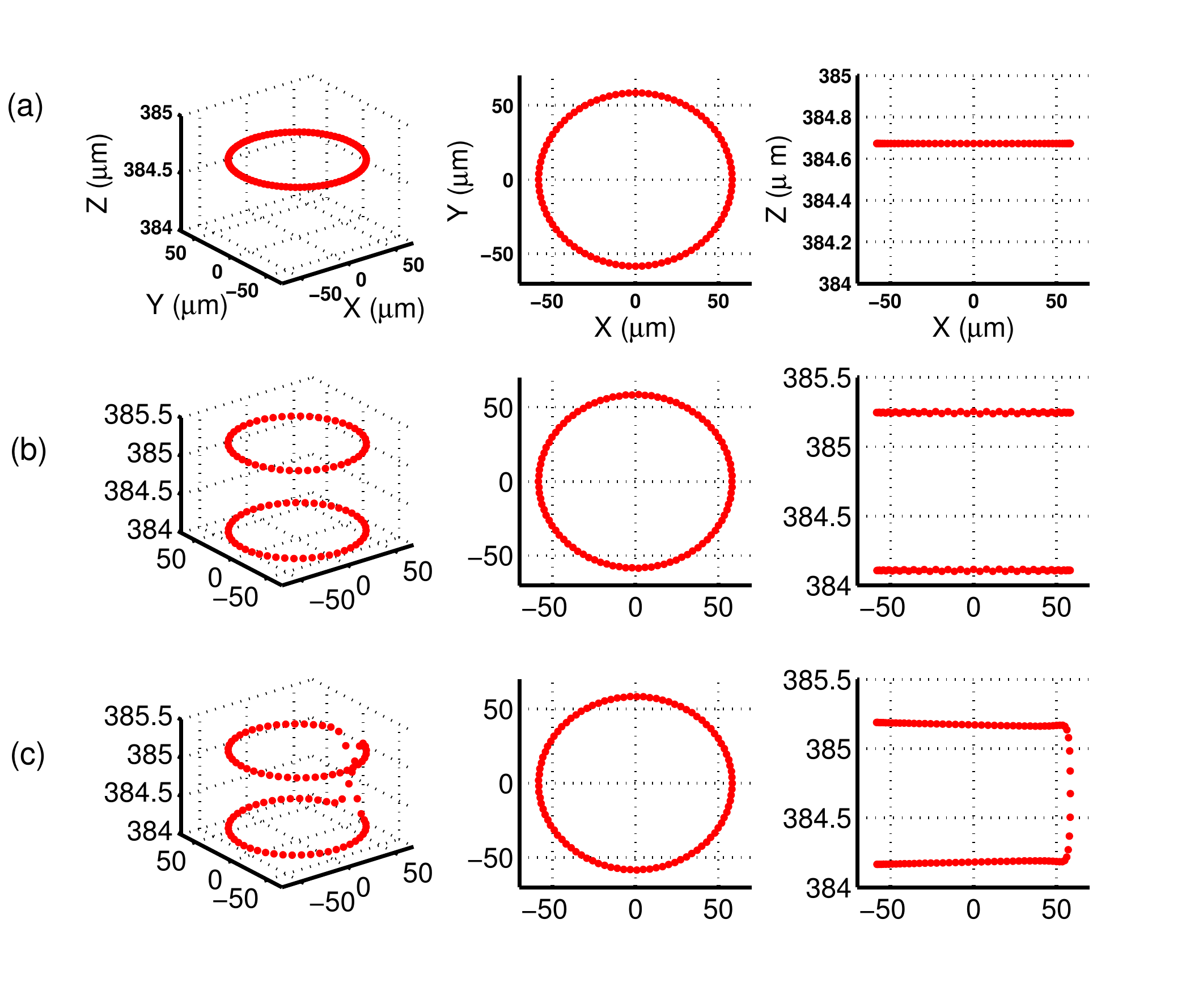}
			\caption{(Color online) Structure of the ring crystal. (a) 92-ion ring 
structure when the radius of center electrode is 126~$\mu$m. (b) 93-ion ring has 
two-ring structure when center electrode is 126~$\mu$m. (c) 93-ion ring also has 
meta-stable kink when center electrode is 126~$\mu$m.}
			\label{fig:crystal}
			\end{center}
			\end{figure*}

		Subsequently, the equation of motion is numerically solved by 
fourth-order Runge-Kutta method with a time step of 20~ns. For a reasonable run-time of the algorithm of a few 
hours, we choose the damping coefficient $\gamma = 2 \times 10^{-18}$~kg/s. As 
a result, our simulation shows that a ring crystal can be formed with up to 
92~ions with the parameters and geometry discussed above (outer radius of the 
inner electrode 126~$\mu$m). As shown in Fig.~\ref{fig:crystal}(a), the 92-ion ring has diameter 116~$\mu$m 
and height 385~$\mu$m, 

Keeping all parameters fixed, but adding one more ion yields a 93-ion ring 
crystal of two layers with about 1 micrometer separation in the plane 
perpendicular to the trap surface, as shown in Fig.~\ref{fig:crystal}(b). 
We can study this phase transition from single-layer ion rings to 
double-layer ion rings by fine-tuning the trapping potential. This can be done 
by adjusting the compensation rf voltage on the second ring, as we have 
discussed in Sec.~\ref{sec:how-to-calculate}. In the double layer regime, we also find 
meta-stable kinks as shown in Fig.~\ref{fig:crystal}(c). The kink dynamics in a 
ring might be an interesting subject in its own right \cite{Landa2010}. Contrary 
to studies in linear traps \cite{Mielenz2013,Ulm2013,Pyka2013}, the kinks are in 
a homogeneous environment and cannot escape by just traveling to the edge of the 
ion crystal. Furthermore, working
with an odd number of ions enforces the presence of an odd number of kinks and 
thus of at least one, while working with an 
even number of ions would lead to an even number of kinks. 		

\section{Analysis of imperfections}
\label{sec:imperfections}
\begin{figure}
			\begin{center}
			\includegraphics[width=3in]{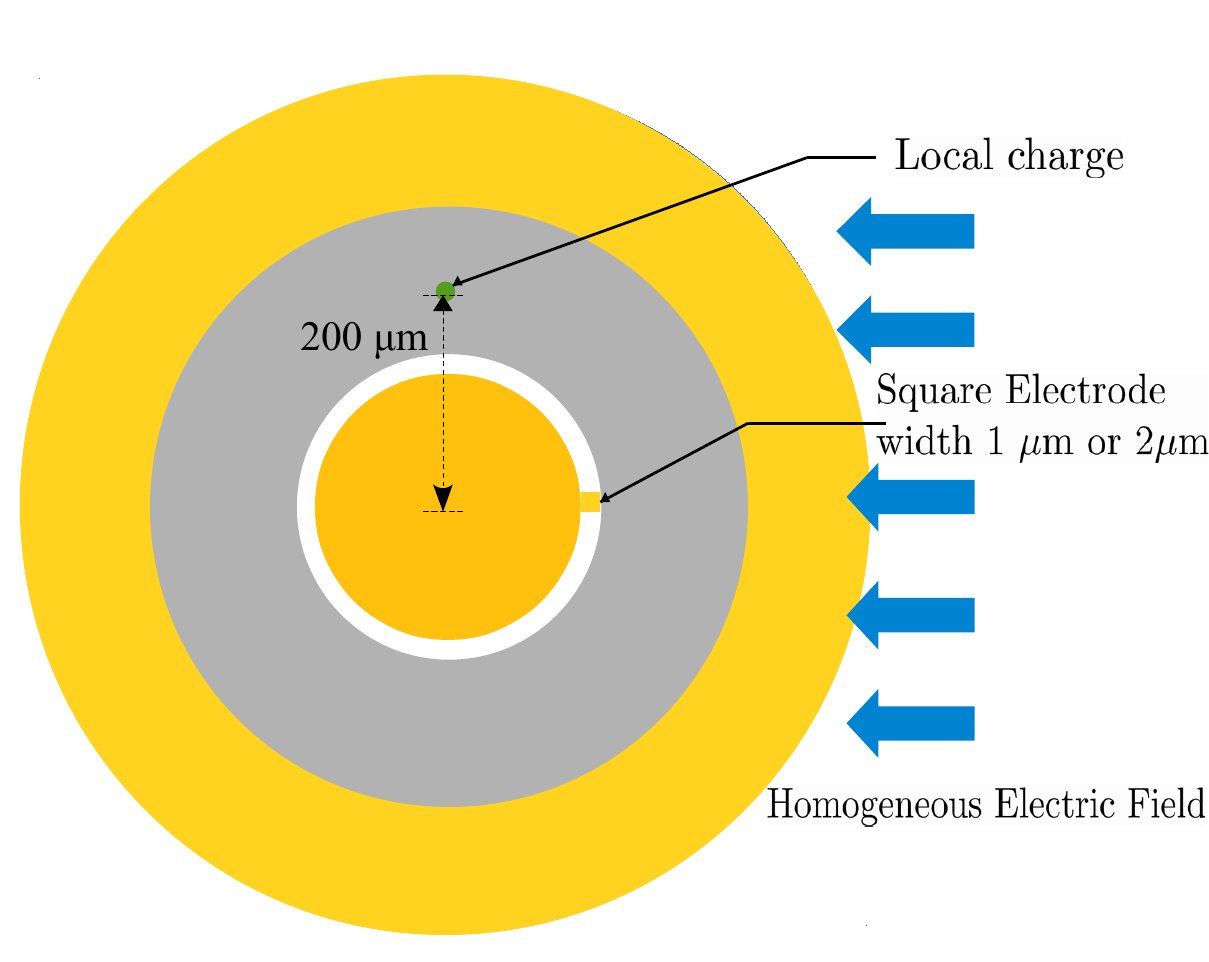}
			\caption{(Color online) Three sources of imperfection: (a) Applying an homogeneous electric field.(b) Adding a 
small square electrode on the edge of the center electrode.(c) Local 
charged dust on the trap surface. }
			\label{fig:imperfection}
			\end{center}
			\end{figure}

	\paragraph*{}
		Of particular interest in our work is to create ion crystals 
 freely rotating even if their
		rotational energy is comparable to the groundstate energy \cite{Li2012,Wilczek2012}. 
With with the criterion established in Eq.~\ref{eq:gap}, {\em i.e} 
the energy barrier created by the imperfections should be be smaller than $E_{\rm 
gap}/2=\sim 0.1$~nK,  we calculate the energy as a 
function of the angle when rotating the crystal around the symmetry axis. We 
study three sources of imperfections: a homogeneous electric field, irregularities on the edge of the electrodes, 
and the effect of a local charging (Fig.~\ref{fig:imperfection}).

\begin{figure}[ht!]
			\begin{center}
			\includegraphics[width=3.5in]{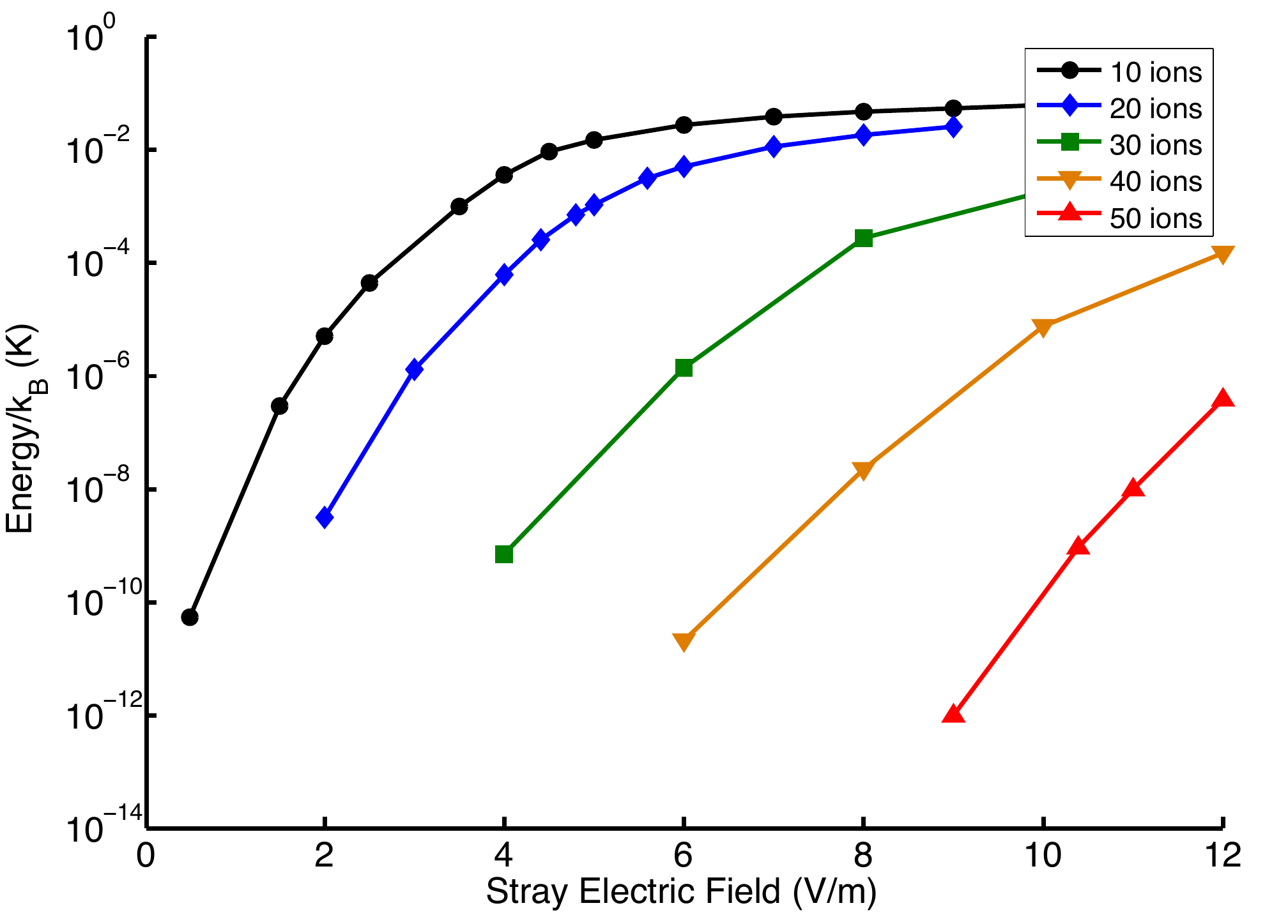}
			\caption{(Color online) Rotational energy barrier as a function of 
homogeneous stray electric field. The barrier drastically reduces with 
increasing number of ions. For 70 ions, the barrier due to homogeneous electric fields can be neglected.}
			\label{fig:stray}
			\end{center}
			\end{figure}

\begin{figure}
			\begin{center}
			\includegraphics[width=3.5in]{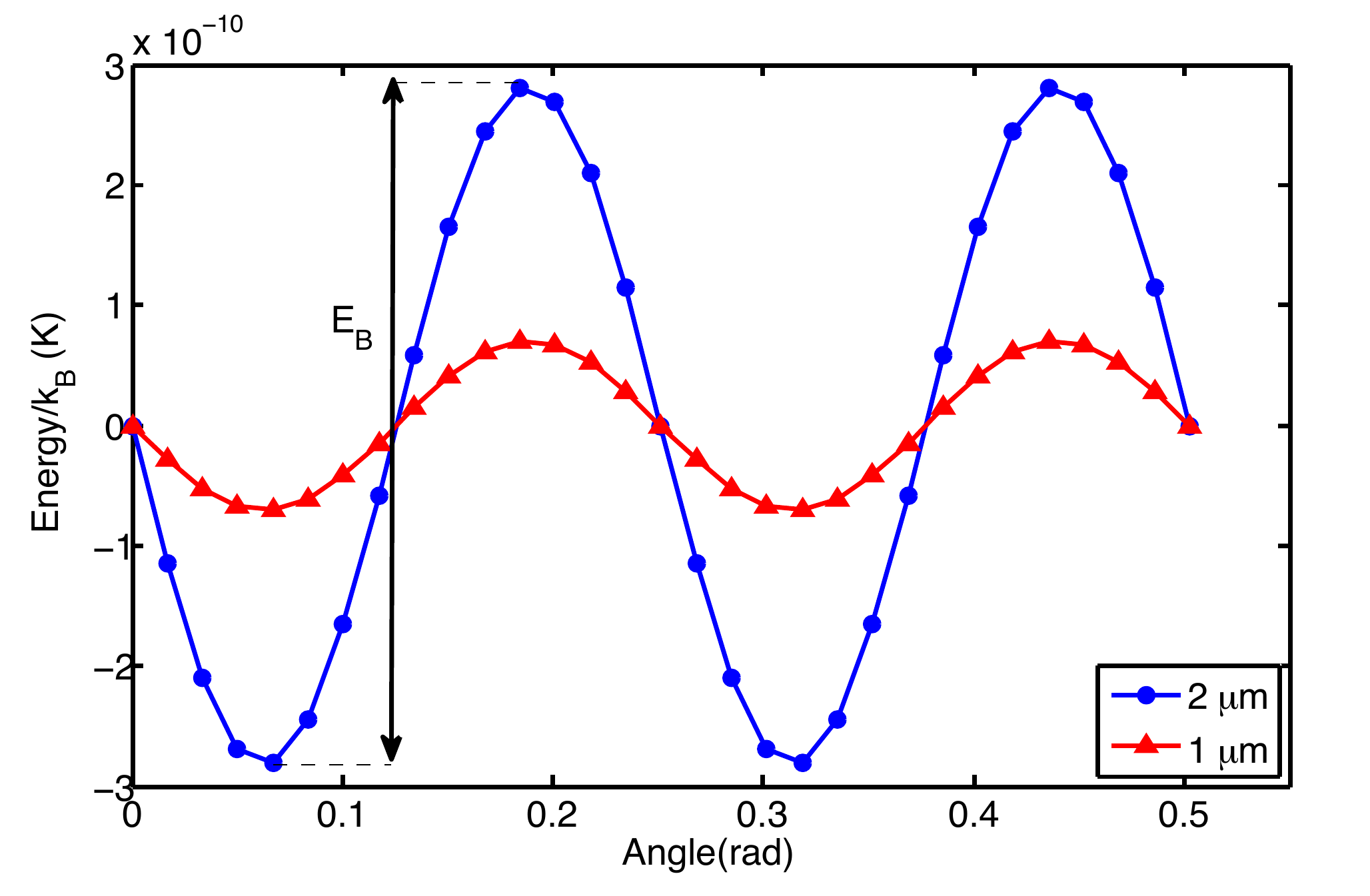}
			\caption{(Color online) Energy as a function of rotation angle of a 
25-ion ring in presence of 1~$\mu$m and 2~$\mu$m square electrode imperfections. 
In analogy to Fig.~\ref{fig:stray}, we expect the barrier to substancially 
shrink for larger ion numbers.}
			\label{fig:square}
			\end{center}
			\end{figure}

\paragraph*{Homogeneous Electric Field} 
		First we calculate the energy of the ion ring as a function of rotation angle in presence of an 
homogeneous electric field. The result will be a sinusoidal periodic function whose amplitude represents the 
classical energy barrier, which we denote $E_{B}$ (c.f.~Fig.~\ref{fig:square}). 
Fig.~\ref{fig:stray} shows the energy barrier $E_B$ 
as a function of the applied field for 10, 20, 30, 40, and 50~ions. 
Fig.~\ref{fig:stray} illustrates that the energy barrier is drastically 
suppressed when the number of ions is increased. Recall that the energy gap 
$E_{{\rm gap}} = N\hbar^2/Md^2$ is also proportional to the number of ions, $N$. 
This suggests that freely rotating crystals with larger ion numbers are easier 
to observe, and suffer less from the imperfection of the trap. Intuitively, this 
can be understood in the following manner: for an increased ion number, the 
ion-ion spacing is much reduced approaching a more homogeneous and continuous 
charge distribution. For such an homogeneous charge distribution external 
imperfections cannot exert a torque on the charge distribution. Thus, the 
rotational barrier caused by the imperfection drastically reduces with an 
increasing number of ions.  
	
\paragraph*{Electrode Edge Irregularity}			
		Next we add a square electrode of width 1~$\mu$m (2~$\mu$m) 
adjacent to the center electrode imitating a fabrication imperfection. The 
electric potential of a square electrode can be calculated from solving 
Laplace's equation analytically \cite{Gotoh1971485}. We obtain the full potential by
superposing this 
solution to the one obtained earlier.  Fig~\ref{fig:square} shows the energy of 
a 25-ion ring as a function of rotation angle with the square electrode of width 
1~$\mu$m and 2~$\mu$m.  
		 We concluded that already for 25 ions, the energy barrier is 
sufficiently small.
Increasing the ion number reduces the ion-ion spacing which is expected to 
reduce the energy barrier further. 
		
\begin{figure}
			\begin{center}
			\includegraphics[width=3.2in]{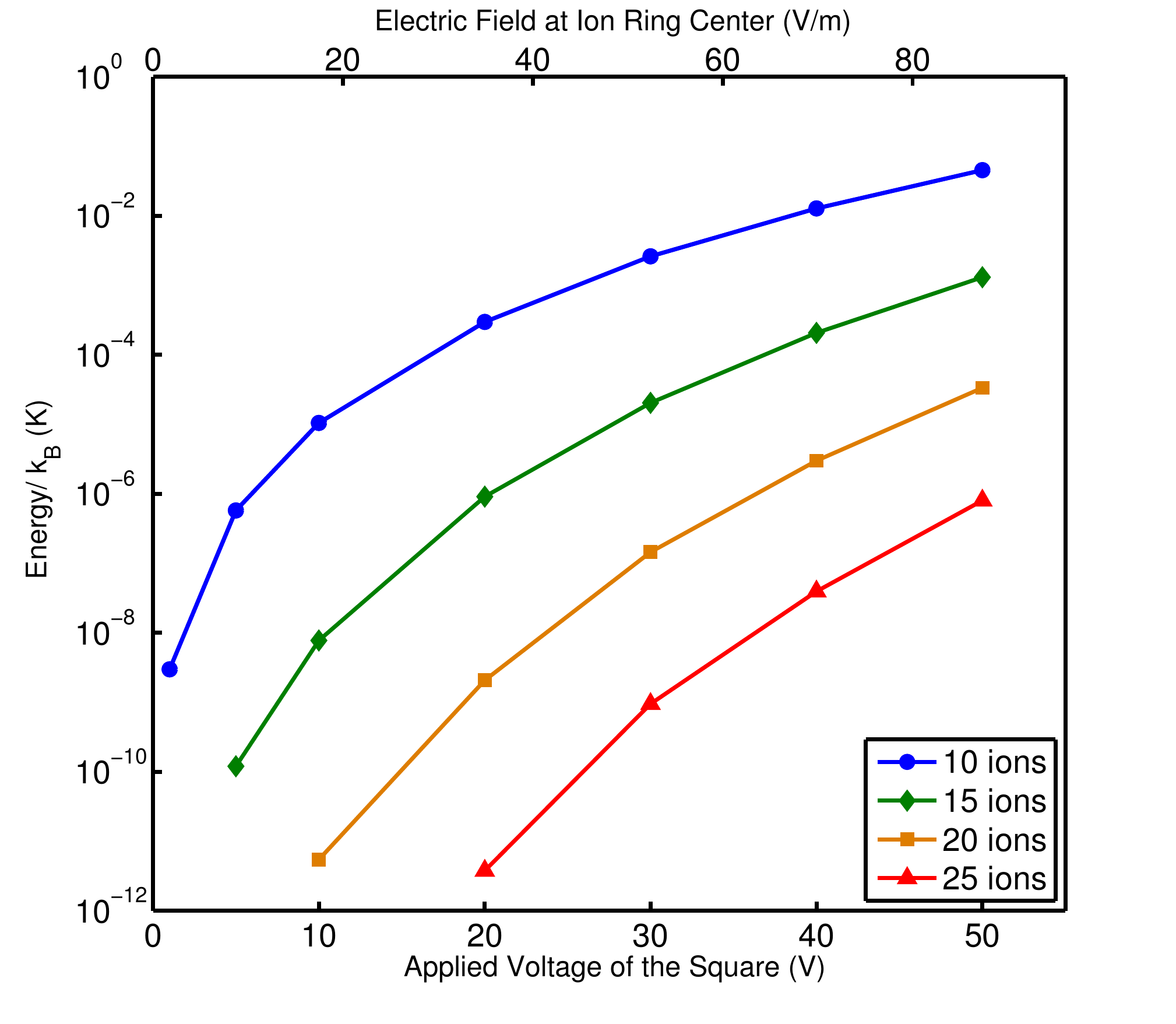}
			\caption{(Color online) Rotational energy barrier as a function of the 
DC voltage on the square electrode (lower x-axis) and its corresponding field strength at the center of the 
ion ring (upper x-axis).  The square electrode of width 10~$\mu$m is
placed on the surface of the electrode and  200~$\mu$m away from the center. This 
simulates charged dust particles in the square that carries a different 
voltage than the rest of the electrodes. Here we calculate the the energy barrier $E_{B}$ of 10, 15, 20, and 25-ion rings. As in
Fig.~\ref{fig:stray}, we expect the barrier to substantially shrink for large 
ion numbers.}
			\label{fig:localcharge}
			\end{center}
			\end{figure}

\paragraph*{Local Charge}
	 In surface traps with ion-electrode distances
on the order of 100~$\mu$m, we typically find electric fields 
on the order of 100~V/m before performing micromotion compensation. 
These fields come potentially from local charges on the trap surface 
as caused for instance by small charged dust particles on the electrodes. 
To study the effects of this, 
we assume that a square of size 10~$\mu$m $\times$ 10~$\mu$m at a position 200~$\mu$m from the center, which carries a different voltage than the rest of the electrodes. By applying 1V to 50 V DC to the square electrode, we create electric
field of strengths from 2 V/m to 100 V/m at the center of the ion ring 385~$\mu$m above the surface. 
The result is 
presented in Fig~\ref{fig:localcharge}. We find that already a 25-ion crystal is nearly insensitive to charging of a 10~$\mu$m $\times$ 10~$\mu$m 
surface to tens of volts. Again, 
increasing the ion number will reduce the energy barrier further, and thus allow for even
stronger local charge imperfections.

\section{Laser Cooling}
\label{sec:lasercooling}
Finally, we study the laser cooling dynamics for freely rotating ion rings. 
In ion trapping experiments, usually a single cooling beam is preferred avoiding 
complications with interference effects. If this beam is
 centered perfectly on the ion ring, the radiation 
pressure will be balanced on each side of the nearly freely rotating ring. If, however, 
the beam is displaced from center, then a net torque from radiation 
pressure will result. Treating the ring as a rigid rotating body, we can apply 
the force equations for Doppler cooling \cite{Leibfried2003} and 
determine the net torque on the ring. In the limit of small velocities, 
at temperatures close to the Doppler cooling limit, we linearize the
forces and the average torque $\mathbf{ \tau_a}$ on 
a single ion can be written as:
\begin{equation}
\mathbf{ \tau_a} = \mathbf{R} \times \mathbf{F_0}(1+\kappa \mathbf{v})\:.
\end{equation}
where $\mathbf{R}$ is a radial vector of each ion. $\mathbf{F_0}$ is averaged radiation pressure in the 
direction of laser beam propagation, 
\begin{equation}
\left | \mathbf{F_0} \right | = \hbar k \Gamma \frac{s/2}{1+s + (2\Delta/\Gamma)^2}\:.
\label{eq:torque}
\end{equation}
The drag coefficient for cooling,
\begin{equation}
\kappa = \frac{8k \frac{\Delta}{\Gamma^2}\cos{\theta}}{1+s + (2\Delta/\Gamma)^2}\:.
\end{equation}
is negative when the detuning $\Delta$ is negative. The $\cos{\theta}$ arises from the 
projection of the laser onto the rotational
degree-of-freedom of each ion, $\delta_{{\rm eff}} = \Delta - k \cdot v$, where $v$ 
is the velocity of the ion. The decay rate is given as 
$\Gamma = 1/(7$~ns) for Calcium, the wavenumber is defined as $k = 2 \pi/(397$ 
nm), and $s$ is the saturation parameter \cite{Leibfried2003}. In what follows, we
assume a saturation parameter of $s=\frac{1}{2}$ and a gaussian beam profile 
with a beam waist of 200~$\mu$m. We chose a large beam waist to minimize the 
differences in intensity across the ring of ions and assume a detuning of $\Delta = 
-10$~MHz for optimal cooling.

The Doppler cooling action of the laser on the moving ions will counterbalance 
the torque from the radiation pressure. If the torque is sufficiently small, 
the forces will cancel at some finite 
frequency of rotation. For 30 ions confined to a $R = 58$ $ \mu$m ring, we 
calculate the equilibrium frequency and find a rotation frequency of 
approximately 1.04~kHz per 1~$\mu$m displacement of the cooling beam. 
\begin{figure}
			\begin{center}
\includegraphics[scale = .4]{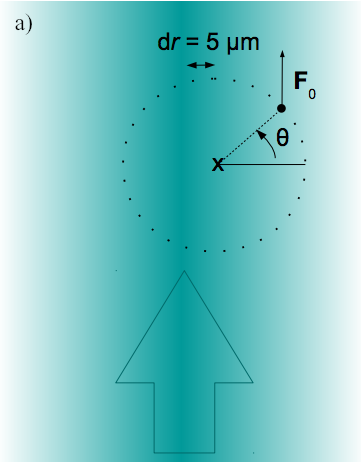}
			\includegraphics[width=3in]{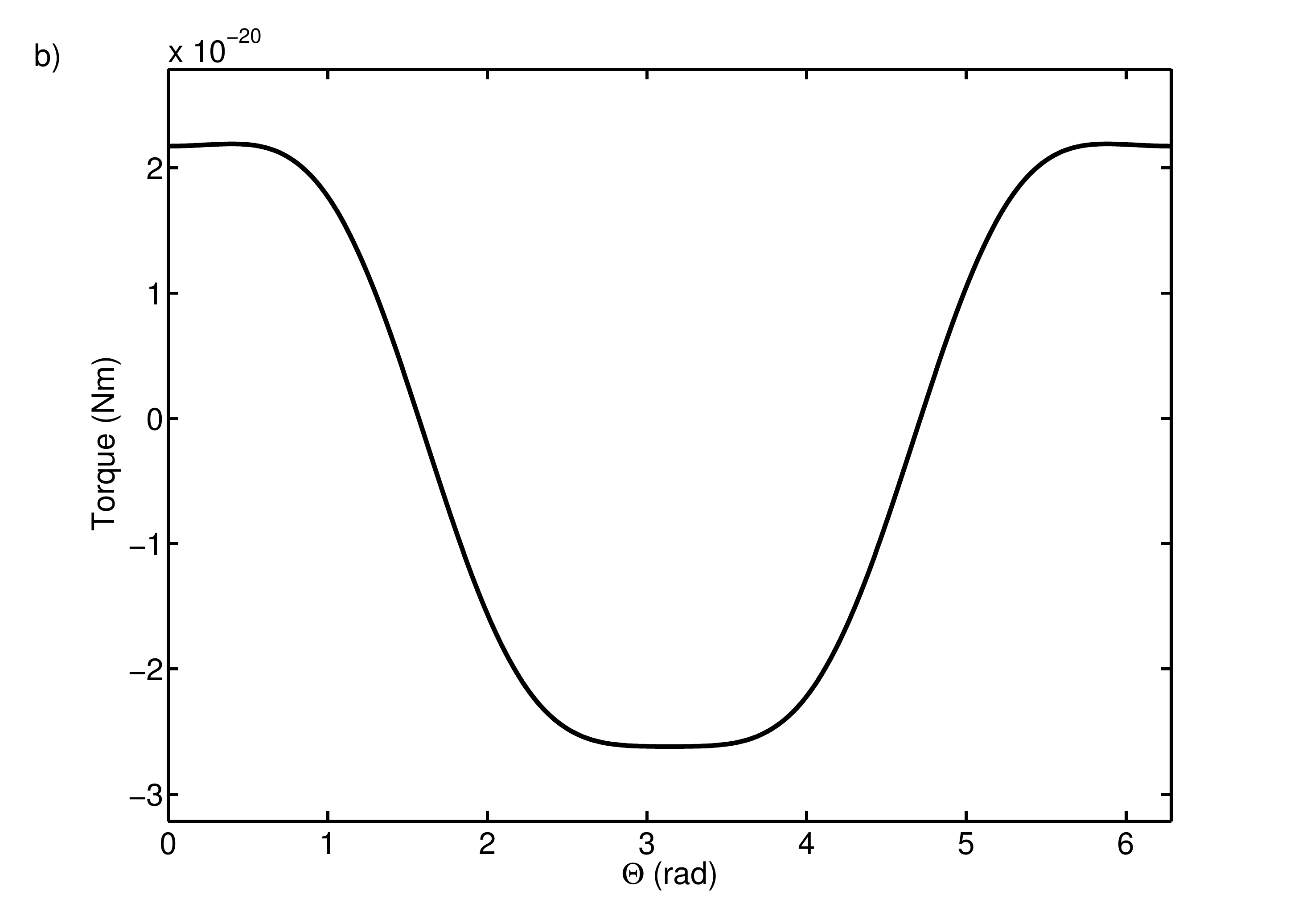}
			\caption{(Color online) (a) Diagram showing the x-y plane at the trapping minimum of the z-direction. The shading indicates the distribution of the Gaussian laser beam, and the large arrow indicates the direction of laser propagation, and therefore the direction of the radiation pressure force, $F_0$. The center of the laser beam is offset with respect to the trapping position (dashed ring) by an amount d$r$. The position of an ion in the ring is defined by $\Theta$ as shown here as and used in the graph, (b). (b) Torque on a stationary single ion due to radiation pressure as a function of position around the ring as described by equation \ref{eq:torque}. We assume a saturation parameter of $s=\frac{1}{2}$, a gaussian beam profile with a beam waist of 200~$\mu$m, a ring diameter of 116~$\mu$m, and a laser displacement of d$r=5 \mu$m.}
			\label{fig:torque}
			\end{center}
\end{figure}

To counteract the rotation, we turn to the previous discussion of the effects of 
electric fields on the ring. By applying a strong electric field, we aim at
creating a sufficiently strong energy barrier, thus stopping the ion ring
from rotating. The maximum 
slope of the energy of the ion crystal as function of the rotation angle 
represents the torque needed to overcome the energy barrier and cause the ions 
to rotate around the ring 
\begin{equation}
\tau_E = \left | \frac{dE}{d\theta} \right |_{{\rm max}}\:.
\end{equation}

Aiming at a static ion crystal, the torque from the cooling laser on a single 
ion is $\tau_L = \mathbf{R} \times \mathbf{F_0}$. Next, 
we look to find an applied electric field that will cancel the torque from the 
laser for this displacement. This condition can be written as 
\begin{equation}
\sum_{i=1}^{N} \tau_L^i < \tau_E\:.
\end{equation}
where $\tau_E$ represents the sum over the torque of all ions by a given electric field, 
with the field orientation chosen to maximize the torque, {c.f.} Fig.~\ref{fig:torque}b).

\begin{figure}[ht!]
			\begin{center}
			\includegraphics[width=3.5in]{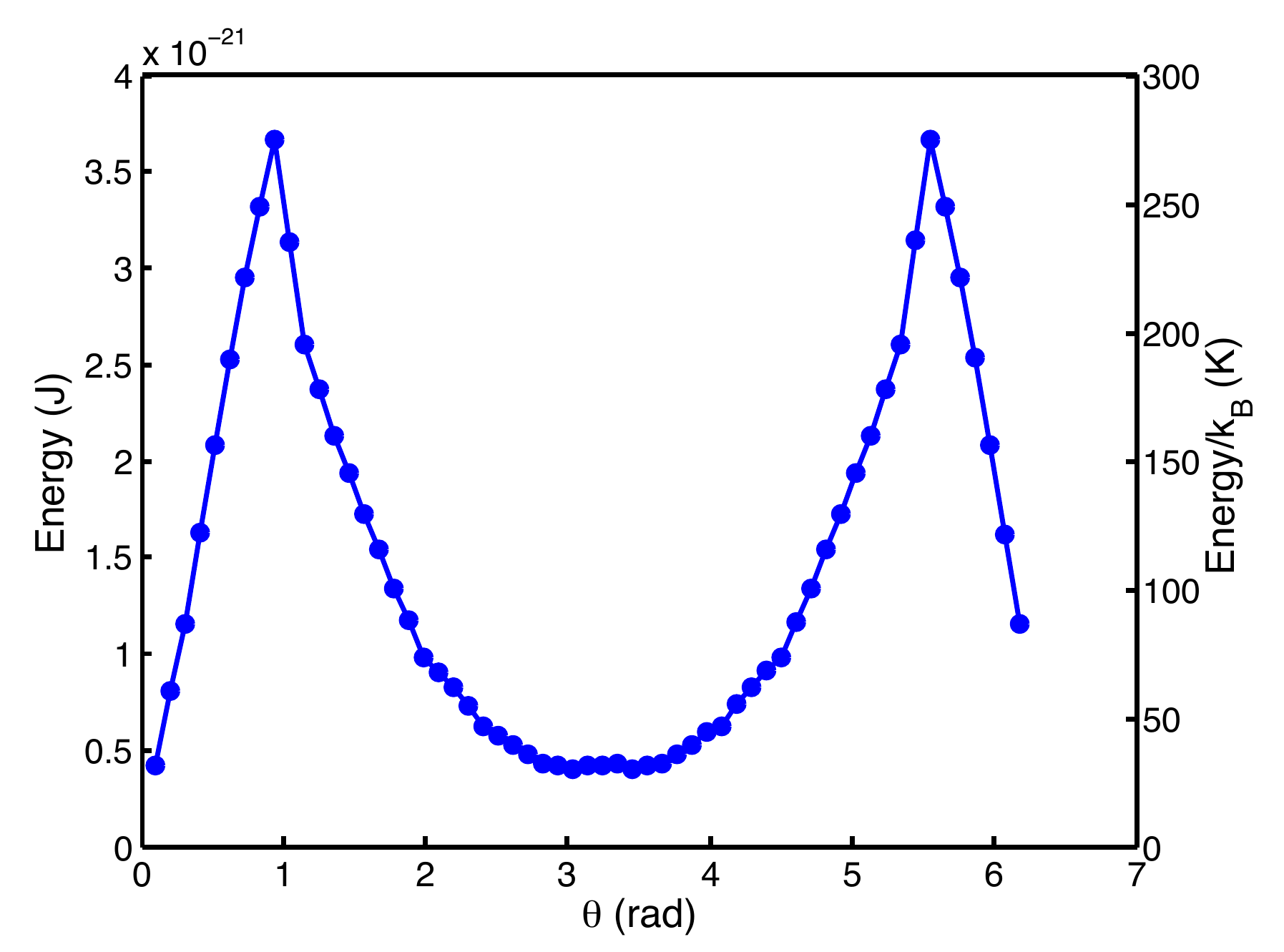}
			\caption{(Color online) Rotational energy of a 30-ion ring under 
homogeneous electric field of 75 V/m. The torque calculated from the maximum 
slope of this curve can counteract the torque from the laser cooling. }
			\label{fig:etorque}
			\end{center}
			\end{figure}

For ion positions in the trap given an electric field of 75~V/m optimally aligned against the propagation of the 
laser, the total sum of the torque from the laser alone is approximately linear for small laser displacements d$r$. Analyzing the situation 
for 30 ions and taking into account their calculated positions in the ion ring, 
we obtain $\tau = -5.476\times 10^{-15}$N$\cdot \textrm{d}r$. 
The torque from the energy barrier for 75~V/m, given by the 
maximum slope of the curve in Figure \ref{fig:etorque},  is $3.3 \times 
10^{-20}$N$\cdot$m, allowing for a displacement d$r$ of up to 6.0~$\mu$m to achieve a 
static ion crystal. 

\section{Conclusions}
	We have studied a design of a planar trap providing trapping of a 
92~ion-ring of diameter 116~$\mu$m at height 385~$\mu$m above its surface. 
This design can be 
fabricated using micro-fabrication methods with high precision. We also studied 
the rotational motion under three symmetry-breaking imperfections: homogeneous 
electric fields,
irregularities of electrode edges from fabrication imperfections, and local charges placed on the trap electrodes.
We have shown that the rotational energy barrier induced by these imperfections 
drastically reduces with an increasing number of ions in the ring. We thereby 
expect that the energy barrier from the imperfections 
can be reduced below the rotational ground state energy of large ion crystals.
In addition, we have shown that laser alignment and strong homogeneous 
electric field of 75~V/m can be utilized to pin and cool the ion ring for 
trapping and imaging.  

\section*{Acknowledgements}
This work is supported by the W.M. Keck Foundation. We acknowledge the contributions of Anthony Ransford and Hao-kun Li to the discussions related to this work.

\bibliographystyle{apsrev4-1.bst}
\bibliography{library}

\end{document}